\documentclass[conference]{IEEEtran}
\IEEEoverridecommandlockouts
\usepackage{balance}
\usepackage{cite}
\usepackage{subfigure}
\usepackage{amsmath,amssymb,amsfonts}
\usepackage{algorithm2e}
\usepackage{graphicx}
\usepackage{textcomp}
\usepackage{xcolor}
\usepackage{array}
\def\BibTeX{{\rm B\kern-.05em{\sc i\kern-.025em b}\kern-.08em
    T\kern-.1667em\lower.7ex\hbox{E}\kern-.125emX}}
\begin{document}
\title{An Integrated P2P Framework for E-Learning
\thanks{Identify applicable funding agency here. If none, delete this.}
}
\author{
\IEEEauthorblockN{Nikita Bhagatkar}
\IEEEauthorblockA{\textit{Department of CSE} \\
\textit{Indian Institute of Technology}\\
Kanpur, India}
\and
\IEEEauthorblockN{Kapil Dolas}
\IEEEauthorblockA{\textit{Departmentt of CSE} \\
\textit{Indian Institute of Technology}\\
Kanpur, India}
\and
\IEEEauthorblockN{R. K. Ghosh}
\IEEEauthorblockA{\textit{Departmentt of CSE} \\
\textit{Indian Institute of Technology}\\
Kanpur, India\\
rkg@cse.iitk.ac.in}
}

\maketitle



\begin{abstract}
The paper describes the design and development of a Peer-to-Peer Presentation System (P2P-PS) that supports live video streaming of lectures coupled with a shareable whiteboard. The participant may resolve doubts during an ongoing session of presentation using {\em ask doubt} feature. It is disseminated to all other peers by preserving the causality between ask doubt and its resolution. A buffered approach was employed for enhancing the performance of the shareable whiteboard which may be used either in tandem with live streaming or in standalone mode. The proposed P2P-PS supported also by a P2P file sharing and searching system with a few additional features. The combined framework allows creation of mash-up presentations with annotations, comments, posts on video, audio and PDF files. The back-end also has a provision for discussion forum. We implemented the P2P file sharing and searching system using de Bruijn graph overlay. Our experiments on P2P-PS was carried out with 1000 peer nodes on Emulab using 200 physical nodes. The results show that the proposed overlay eventually stabilizes even in the presence of a churning rate of up to 30\%. The maximum path length being just six hops. The estimated throughput is found to be close to our theoretical results. The experiments on de Bruijn graph overlay reported an average out-degree of 7.99 per a node in a graph consisting of 800000 nodes. Only less than $0.4\%$ nodes have out-degrees greater than $16$. The minimum and the maximum in-degree were $7$ and $8$ respectively. These results match with the known theoretical results on the analysis of de Bruijn graphs. The Emulab experiments were also carried out for determining the maximum latency and the success rate for look-ups in the presence of churning. The results show that the look-up success rate is 99.39\% even when nodes leave the system after every three minutes with 10\% probability. 
\end{abstract}

\section{Introduction}

The cloud-based collaboration platforms are of three types, namely,
\begin{enumerate}
    \item Collaborative creation of forms, reports or documents (e.g. Overleaf~\cite{overleaf:live}, Google Drive~\cite{google:live}).
    \item Collaborative development of large computer programs (e.g. Github~\cite{github:live}).
    \item Peer to peer tutoring/reciprocal learning by teaching (e.g, Duolingo~\cite{duolingo}, Coursera~\cite{coursera}, Kahoot~\cite{kahoot}, Brainly~\cite{brainly}). 
\end{enumerate}
These collaboration tools are inspired by the concept of social networking. The underlying idea is to combine online access to web-based interfaces that can some what mimic classroom lectures with a Facebook like secure interaction environment between the teachers and the students. A few of these collaboration tools include online tests, assignments or links to external web materials. In summary, social learning platforms provide a framework for a coordinated learning environment or a discussion forum.

Brainly~\cite{brainly} follows a slightly different approach. It harnesses crowdsourcing of like-minded students to combine their problem-solving skills. It uses deep learning techniques based on historical data to predict special requirements of a learner. A student gets a personalized learning experience when he/she gets connected to an appropriate peer-mentor who helps the student in understanding the problem or develop the required skills. 

However, the main focus of any social learning platform is essentially a {\em passive discussion forum} with variations such as crowdsourcing or incorporating additional features like peer mentoring. The elements of an active learning as it happens in a live classroom or a physical meeting can be captured by combining a P2P Presentation System (P2P-PS) with the features of an video conferencing system such as Skype~\cite{baset2004} or Hangout~\cite{icard2014}. The major difficulty in the suggested approach is scalability, e.g., group call limit in Skype is only 25. Most video conference systems include screen sharing feature. However it is one-to-many as many-to-many sharing does not make sense in the context of a conferencing system. Our approach, therefore, is novel as it incorporates live video streaming aided by a shareable whiteboard. We propose an immersive solution by coupling the P2P-PS with a P2P file sharing an searching system at the back-end. The back-end depends on a conventional Chord-like DHT~\cite{stoica2003chord} but uses a de Bruijn graph~\cite{} based implementation of DHT for file sharing, file searching. It provides a number of features such annotations, comment and discussion forum. So, one can create a mashable presentation.  

The rest of the paper is organized into six sections.  There are two possible organizations of peers for live streaming, namely, (i) tree and (ii) mesh. Section~II deals with the inadequacies associated with the tree organization and then explains why a pure mesh organization is inappropriate for the live streaming in an implementation of our system due to an additional feature which allows non-generating peers to seek clarifications during the streaming. Section~III describes the modified form of mesh organization which allows us to exploit the availability of spare capacities at peers so that dynamic data-driven fanouts at peers can support live streaming. Joining and leaving of peers, handling of 'ask doubt' feature and memory utilization by modified mesh organization all discussed in this section.  Section~IV focuses on design and development of Live shareable whiteboard.  The results of experiments on Emulab are discussed in section~V and section~VI concludes the paper.

\section{Live Streaming}
There are two approaches for live media streaming on a P2P system, viz. 
\begin{itemize}
    \item Tree-based: It maintains a logical tree overlay where the root generates the stream, and other peer nodes get it from their respective parents. The leaves are free-riders receiving the flows from the others.
    \item Mesh-based: A mesh architecture allows each peer to connect to other peers. It also lets a peer combine sub-streams received from more than one peer. 
\end{itemize} 
Other live media streaming techniques are variations of the two basic approaches.

\subsection{Inadequacy of Tree-based Live Streaming Approach}
The tree model follows a push-based technique. The data propagates level by level. A node may have multiple children but one parent. If the parent leaves a session, the entire sub-tree below it gets disconnected from the network. Hence, an internal node cannot leave the tree even if it is not interested in receiving the streaming contents any longer. 

Another major disadvantage of a tree is that it cannot handle dynamicity in the network. A peer may leave the tree either gracefully or involuntarily. In either case, all the existing children of the leaving peer would have to search for a new parent. It results in interruption and loss of data streams. The entire sub-tree under the leaving peer would suffer data loss until all the orphaned children can find a new parent. 

The depth of a tree decreases with increase in fanout. Constructing a bushy tree helps in reducing the latency. However, the fanout bound is determined by the upload bandwidth at a node. A few internal nodes may be available with spare capacity. Such internal nodes can accept additional requests for adopting new children. Reconstruction of the tree will not be required if the number of available peers with spare capacity is equal to the degree of the departed node. Otherwise, the entire sub-tree should be reconstructed. The reconstruction of a tree incurs a significant delay in the streaming process.
One of the examples of single tree overlay architecture is SpreadIt~\cite{deshpande2001streaming}.
It uses a single tree for media streaming. But, as stated earlier, it is vulnerable to the dynamic nature of nodes.

The learning platform proposed in this paper supports a feature for seeking clarifications called "{\em Ask doubt}." A peer can invoke it seeking clarification at any point of the live streaming. The question is propagated to the presenter as well as to the other participating peers. The simplest way to implement this feature in the tree model would be to send the question directly to the root peer (the initiator or the speaker of the session). The speaker would then take the responsibility of propagating the doubt to all the descendants in the tree. It guarantees a causality relationship in 'asking a question' and 'resolving it.' Considering the inadequacy of tree-based streaming, we used a mesh-based approach for supporting live streaming.
 
\subsection{Mesh Approach for Live Streaming}

In the mesh-based P2P system, the nodes randomly connect forming an overlay mesh. These connections could be used to deliver data in unidirectional or in a bidirectional way. For example, CoolStreaming~\cite{zhang2005coolstreaming} maintains bidirectional connections while on the other hand, PRIME~\cite{magharei2009prime} maintains unidirectional connections. 
The mesh-based approach uses swarming content delivery mechanism~\cite{stutzbach2005scalability}.  A peer who has received data from a server could itself act as a server for other peers. Swarming is commonly used in BitTorrent~\cite{magharei2007mesh}. A peer collects data parallelly from other peers and combines it in a single file thereby efficiently utilizing the bandwidth of its neighboring peers. It reduces the load on the primary server because many peers now share the distribution content.

\section{Modified Mesh Approach}

We use an approach slightly different from the usual mesh. It creates a mesh based on the availability of spare capacity at a peer. The modified mesh approach is thus data-driven and supports a dynamic fanout. However, the fanout value cannot be unconstrained.  The upload bandwidth and streaming rate determines the fanout. However,  the feeder node of fanout cannot support a faster outflow rate than its inflow rate. A relationship should exist between the in-degree and out-degree of a node for receiving the data packets continuously. We would continue to abuse the terms "parent" and "children" of a node while referring respectively to a neighbor on an inflow path and a neighbor on an outflow path. 

During a streaming session, all the nodes except for the ones connected directly to the source must maintain the relationship, 'in-degree $\leq$ out-degree.'  It does not preclude the relationship 'in-degree $>$ out-degree.' So, a node may have a higher number of inflows than the number of outflows. It essentially implies that a node has more parents than the number of children. However, we discard such a case, because if a recipient gets more packets than it could send then eventually it will run into a buffer overflow problem leading to loss of multiple packets and re-transmissions.

\begin{figure}[htb]
    \begin{center}
         \includegraphics[scale=0.4]{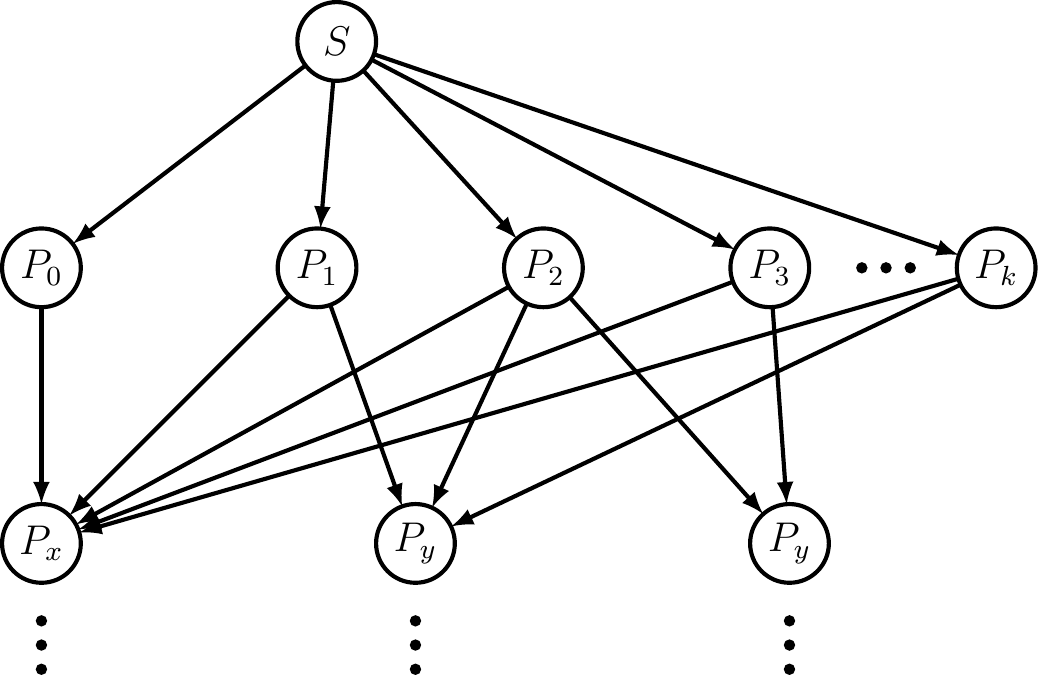}
    \end{center}
     \caption{Partial view of mesh overlay.}
     \label{fig:Mesh}
\end{figure}

\begin{figure}[htb]
    \begin{center}
\includegraphics[scale=0.4]{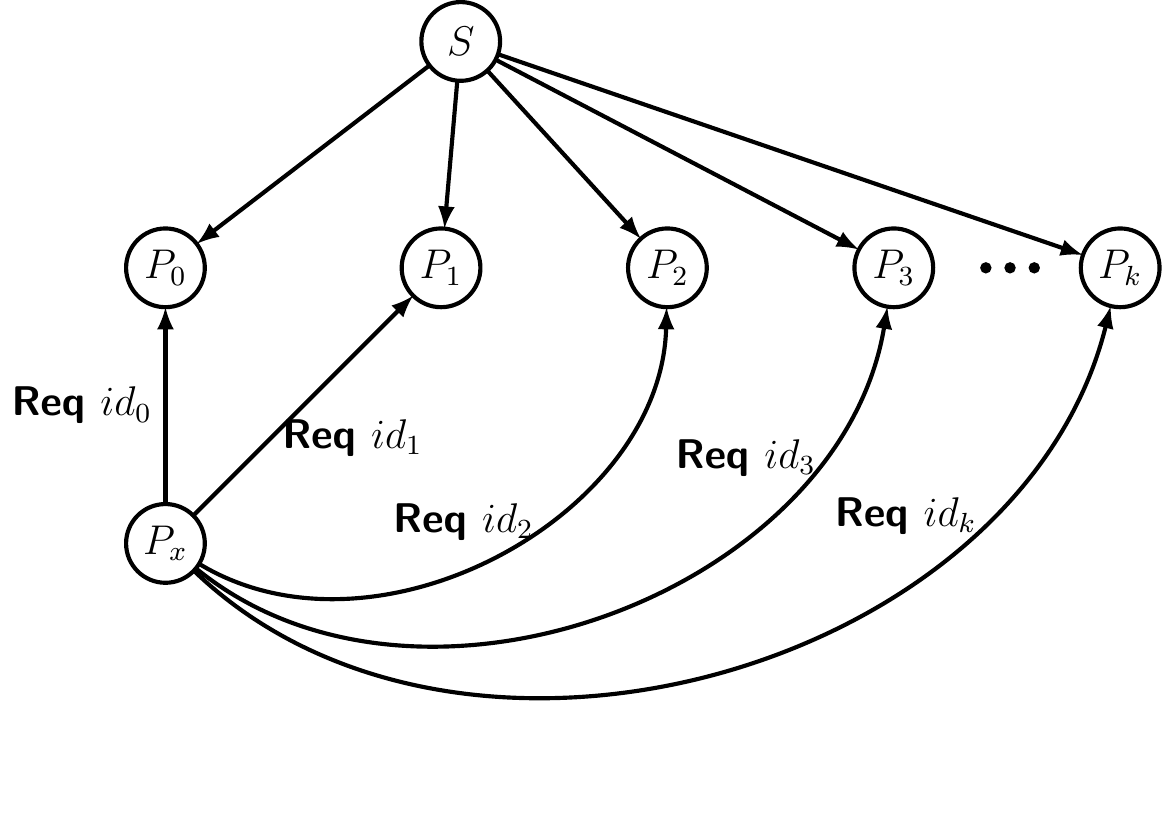}
    \end{center}
    \caption{Initial request.}
    \label{fig:preq}
\end{figure}

\begin{figure}[htb]
    \begin{center}
\includegraphics[scale=0.4]{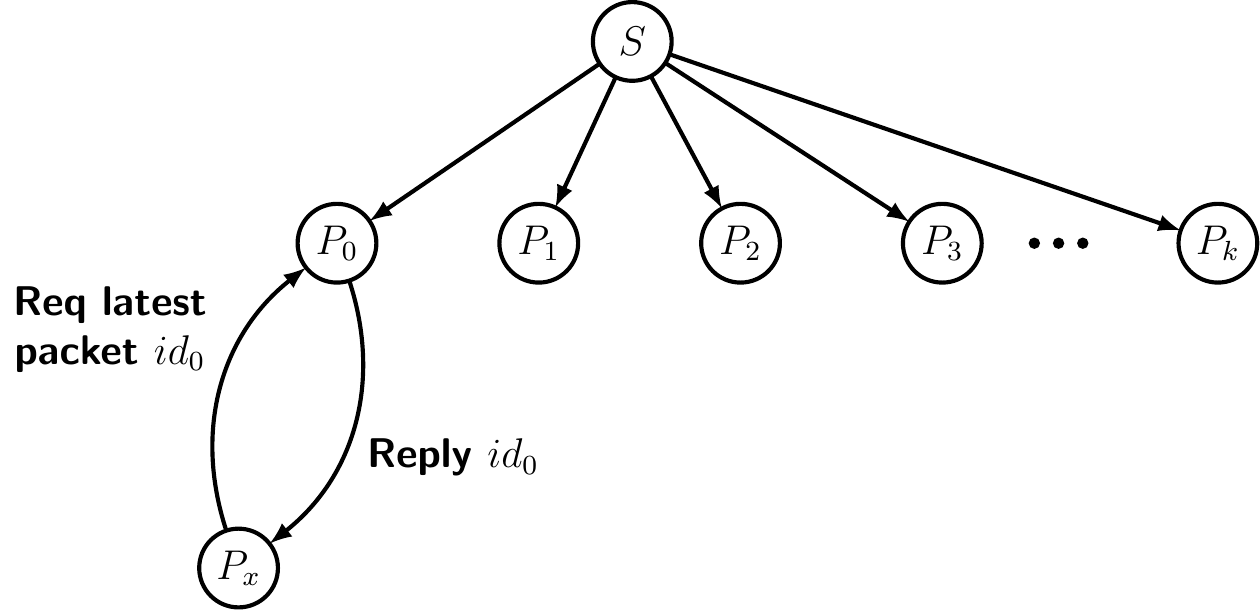}
    \end{center}
    \caption{Request reply sequence for latest packet.}
    \label{fig:repreq}
\end{figure}

\begin{figure}[htb]
    \begin{center}
\includegraphics[scale=0.4]{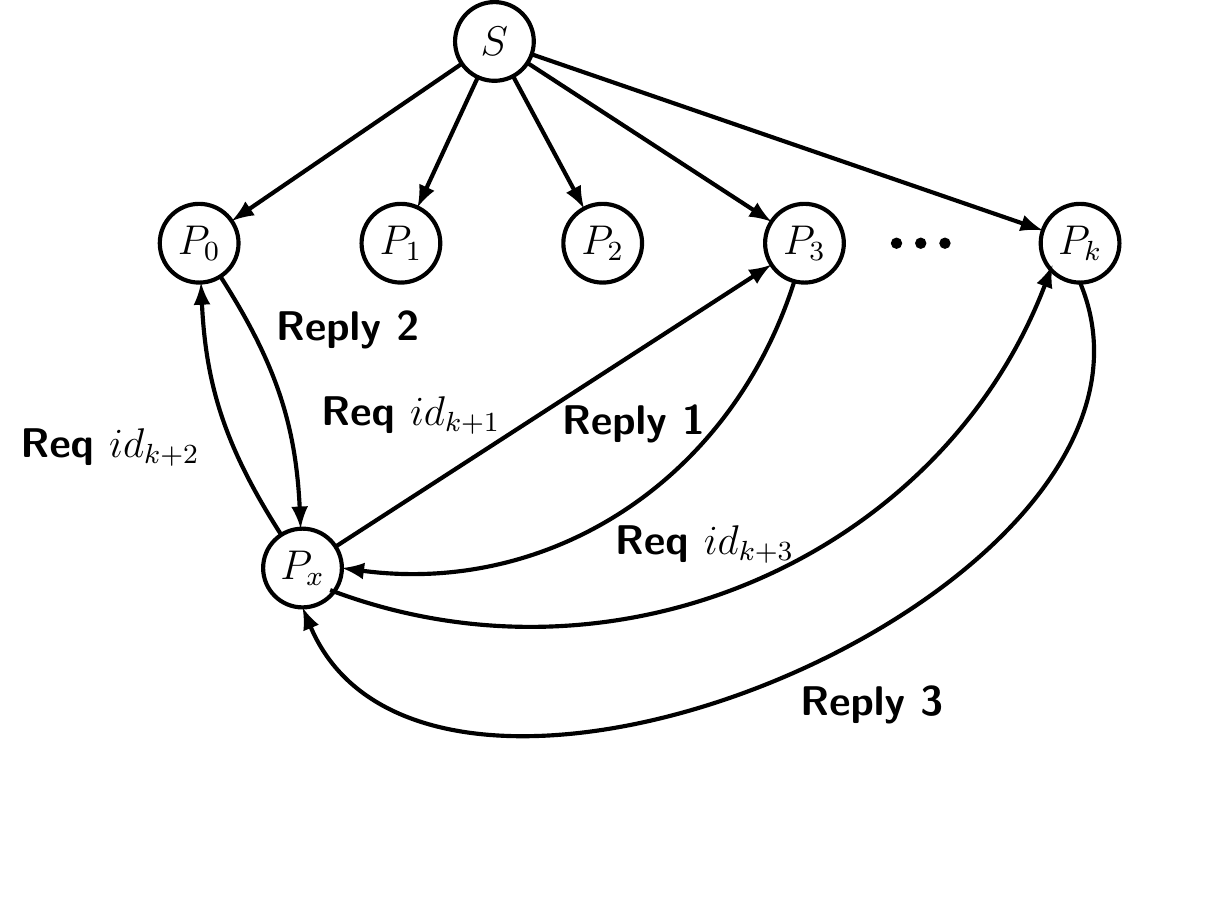}
    \end{center}
    \caption{Request for subsequent packets.}
    \label{fig:subreq}
\end{figure}

We divide the media contents for streaming into chunks. These chunks are propagated to the peers where they could assemble the packets to a media file. The presenter is at the source node which streams the media contents to its neighboring peers. Being a multi-parent, multi-child architecture, a peer could ask for chunks from its parent peers and deliver these to its children. As stated earlier, whenever a source node joins the system, it authenticates itself from the bootstrap server. After authentication, the presenter could start streaming data. However, it will not send data to any of its children unless explicitly asked. Whenever a new peer joins the system, it will get a list of nodes from the bootstrap server.   As soon as the node gets attached to its first parent, it will ask the parent about the latest (current) streaming packet id. Assume that a peer $P_x$  have parents $P_0$, $P_1, P_2, P_3, \ldots, P_k$. Let $P_x$ receive its latest packet information from parent $P_0$. Let the chunks ids are assigned in increasing or $id_0$, $id_1, id_2, id_3, id_4, \ldots$, and so on. Further, let the latest packet be $id_0$. Hence, in round 1, peer $P$ will request for $id_0$ from $P_0$, $id_1$ from $P_1$, $id_2$ from $P_2$, and so on, $id_k$ from $P_k$. Suppose, $P_x$ gets initial reply from $P_3$, then from $P_1$, then from $P_k$, etc. So, peer $P_x$ will ask for $id_{k+1}$ from $P_3$, $id_{k+2}$ from $P_1$, $id_{k+3}$ from $P_k$ and so on. In other words, a peer will ask for the next id from the parent which replies first. It will create a stack of ids at the parent side. This process works parallelly. The above procedure is depicted below in a series of diagrams in Fig.~\ref{fig:Mesh}-\ref{fig:subreq}. Figure~\ref{fig:Mesh} shows a partial view of a random mesh structure during execution of the algorithm. Consider the node $P_x$ and its parents only. 
The initial requests for the latest packets are made to the parent in order from $P_0, P_1, \ldots, P_k$ as depicted in Fig.~\ref{fig:preq}. 
There is no need for a separate scheduler process.  Each peer issues a request for the latest packet in a way akin to a reservation system. 
Now, suppose, $P_x$ gets packet id0 from $P_0$ as indicated by Fig.~\ref{fig:repreq}. 
Figure~\ref{fig:subreq} depicts the process of issuing requests for pulling out subsequent packets after the parent peers reply to the previous requests placed by $P_x$. 
    
        
Whenver $P_x$ receives a packet from any of its parents; it updates the id for the next request.  The process of updating id is performed in a mutually exclusive manner. It could be observed that the request for id is sent to a parent even when the id is not available. It is like making a reservation for a packet of a particular id. With this approach, there is no requirement for having a separate scheduling technique. 

\subsection{Peer Joining}

Initially, the root starts an active session of the mesh overlay. The root is synonymous to the presenter or the teacher. It is responsible for generating the streaming content. An ordinary peer has to authenticate itself before being is allowed to join the overlay. The session key serves as a means of authentication. A peer could join a session only if the root has already started the session. The bootstrap server doubles up as an authentication server in our implementation. Each peer (other than initiator) is authenticated from the session key it submits. The bootstrap server adds the requesting peer to the active peer list after successful. The active peer list consists of all the active peers in the system having spare capacity. The bootstrap server then returns a subset of active peers to the requesting peer.  Let the total number of active peers be denoted by $n$. A subset of $\log_{2}n$ nodes selected randomly from the active peers which joined the system before the newly admitted peer. The reason for choosing this parameter is to decrease the distance between the source and the newly joined peer. It results in improved startup time. Once the list is returned by the bootstrap server, the peer send 'Adopt me' request to all the active peers in the list. A peer could accept or discard the request depending on its fanout value. Fanout of a node is its out-degree. Fanout of a node is determined by:
\[\frac{\text{upload bandwidth}}{\text{streaming rate}}.\]
Hence, the fanout value of each peer is a dynamic value. Initially, we maintain the condition that in-degree = out-degree except for the nodes which are directly connected to the source node. Hence, if the number of parents (in-degree) of any nodes becomes less than its out-degree, then it requests the bootstrap server to send (in-degree - current\_parent\_count) number of active peers. The bootstrap server randomly selects a subset of active peers from the entire active peer list and returns it to the requesting node. 

\subsection{Peer Leaving}

When a node quits gracefully, it proactively informs the other children and the bootstrap server about its exit. However, involuntarily exit of a node is a bit difficult to handle. If the exiting node is not serving any other node at the point of leaving, then there is no disruption in the network. So, let us consider the leaving of a node which serves other nodes. At any point of time, a peer has to remember only the id which it has requested to its parents. Hence, if a parent fails abruptly, it will result in the loss of just one chunk. Assume that it takes $t$ units of time for transferring a chunk between a sender and a receiver.  If there is no response for the $2t$ time, then the peer compares the 'requested id' with the id of the last packet it has received. If the requested id is less than the last packet id it has received, then requester assumes that the server node is either dead or does not have the packet. So, it creates another request id for the same packet to a different peer. It implies that unavailability of one chunk may lead to a delay of $2t+t$=$3t$ time units. The missing chunk is received from one of the other parent nodes. 

If a peer $n$ experience a delay of $\text{time out} > 2t$ units for a response from any of the parent $p(n)$, then $n$ reports $p(n)$ to the bootstrap server.  The bootstrap server then initiates 'are you alive?' message to $p(n)$. If there is no response for a time exceeding $2t$, then $p(n)$ is assumed to be dead. The bootstrap server removes the non-responding node from the list of active peers list and informs the child node $n$ about the loss of its parent so that it could update its parent list. However, no parent list exists. A parent store the list its children. Hence, if a node exits involuntarily, then none of its children gets any information about the parent's exit. Every child node has to figure out on its own about the failure of its parent. 

Suppose a node is left with only one parent who also fails abruptly. The orphaned peer must then request the bootstrap server to assign a new set of parents. It might lead to a bit of delay. We used a scheme of proactive allocation of parents to reduce the delay for allocation of parents to the orphaned nodes. Once, a child notices that it has less than $k/2$ parents it requests the bootstrap server to allocate at least one extra parent. Seeking one additional parent is to restrict the effect a {\em flash exit} which occurs when a presentation (lecture) is about to end. Many peers start leaving the system at once. The count of the parents for each existing peer falls below $k/2$ at a faster rate. Therefore, at that point, it might not be possible to satisfy the requirement of $k$ parents. In the worst case, only a source node and a single peer may be available, and all other nodes may have departed. In this case, the source could be the only parent of the remaining peer.


\subsection{Handling "Ask Doubt" Feature}
Every peer is aware of the source node’s (presenter's device)  address.  Whenever a participant of the audience wants to initiate a query, he/she clicks on ’ask doubt’ button provided in the UI of our application. It establishes a direct connection with the presenter's device. The presenter gets a notification of the query and may send an acknowledgment. The peer device is allowed to send the query after receiving the acknowledgment. The doubt or the query should be in the form of  an audio message like it happens during a physical presentation. The message is unicast  on the link between the source and the peer initiated the query.  The child peers of the requester would then pull the data while the source pushes the data to its other children. Thus, the query is propagated in a push-based manner as happens in the tree-based approach. It guarantees that no query is resolved by the speaker before it is asked, thereby preserving the causality relationship.

\subsection{Memory Utilization}
A video packet consists of three key fields: Video data, the start of data, the length of the data. The system maintains a dictionary called PList for storing packets. It stores the data of recent $10\times1024$ packets. Each packet size is determined by an MTU of 1400 bytes. The streaming rate is assumed to be 2 Mbps. 
\begin{itemize}
    \item The number of packets generated per second = $\frac{2\times10^6}{1400\times8}$ = 179 packets.
    \item The maximum amount of video data that can reside in PList is equal to $10\times1024\times1400\approx14MB$. When size(PList)$>=10\times1024$, then for inserting a new packet in the data list, the oldest packet video data is removed from the PList.
    \item The video data stored in PList could run a session for $\frac{10\times1024}{179}\approx 57$ seconds. 
    \item Older data may be needed only if session exceeds this limit. Note that only video data is removed, start and length fields remains unchanged.
\end{itemize} 
Consider the case where session is running for 8 continuous hours. The number of packets generated in 8 hours = $179\times8\times60\times60$ packets. The total memory consumed by PList in 8 hours = $179\times8\times60\times60\times(8+8)+1024\times10\times1400 \approx 100MB$.

\section{Live Shareable Whiteboard}
\label{chap:live-board}
A video packet consists of three key fields: Video data, the start of data, the length of the data. The system maintains a dictionary called PList for storing packets. It stores the data of recent $10\times1024$ packets. Each packet uses an MTU of 1400B. We have used a streaming rate of 2 Mbps. 
\begin{itemize}
    \item The number of packets generated per second = $\frac{2\times10^6}{1400\times8}$ = 179 packets.
    \item The maximum amount of video data that can reside in PList is equal to $10\times1024\times1400\approx14MB$. When size(PList)$>=10\times1024$, then for inserting a new packet in the data list, the oldest packet video data is removed from the PList.
    \item The video data stored in PList could run a session for $\frac{10\times1024}{179}\approx 57$ seconds. 
    \item Older data may be needed only if session exceeds this limit. Note that only video data is removed, start and length fields remains unchanged.
\end{itemize} 
Consider the case where session is running for 8 continuous hours. The number of packets generated in 8 hours = $179\times8\times60\times60$ packets. The total memory consumed by PList in 8 hours = $179\times8\times60\times60\times(8+8)+1024\times10\times1400 \approx 100MB$.

\subsection{Design and Implementation}

The live media streaming uses the pull-based approach to get the data stream packets from the parent peer. However, live shareable whiteboard follows a push-based approach for data propagation.  There is no parent-child relationship between the nodes. All adjacent nodes are neighbors. 
\begin{figure}
    \centering
    \subfigure[Packet ID.]{\label{fig:packet-id}\includegraphics[scale=0.5]{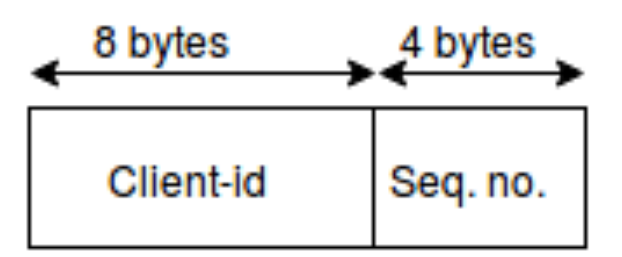}}\hspace{1.5cm}
      \subfigure[Packet structure.]{\label{fig:packet}\includegraphics[scale=0.5]{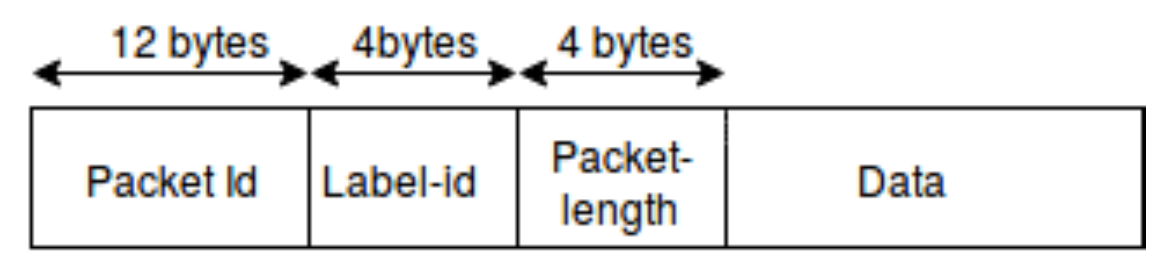}}
    \caption{White board packet.}
    \label{fig:wbpacket}
\end{figure}

A whiteboard packet consists of a packet-id, a label-id, the packet length and the data. The packet-id is composed of the client-id and the sequence number. The structure of a live board packet is shown in Fig.~\ref{fig:wbpacket}. 

The packet-id field uniquely identifies a packet. The client-id is the first 8B of the SHA-1 hash of the IP address of a user.  The client-id ensures that a packet is not delivered to the same peer from whom it is received. Multiple pages are supported in the shareable board. Each page is identified with a unique label-id. The operations performed on every page are stored separately. The label-id  makes the canvas repainting task easier on the receiver side. It also allows maintaining a consistent view of both the sender and the receiver. Due to push-based approach, there may be data redundancy at the receiver side. So, the packets which are already seen are marked and discarded.  Propagating text data is not an intensive operation. Therefore, even with the push-based approach, it does not contribute much to control overhead. 

\subsection{Repainting Live Shareable Whiteboard}

The user interface of the live board application provides many operations like free-pen, basic shapes, color, etc. Among these, free-pen is the most time consuming  A single free-pen operation generates a lot of events. Sending one event over the network consumes a lot of bandwidth and degrades the system performance under the presence of moderate to heavy network traffic. To improve the performance, we used a buffered approach. A bunch of events is stored in a buffer.  The entire buffer contents are sent if either the buffer is full or a mouse release event occurs. The operations are replayed at the receiver's side. It is guaranteed that the events are repainted in a sequential manner only. The size of the buffer could be decided based on the packet size or MTU.

\subsection{Consistency of Board Views at Peers}
Maintaining a consistent view of all peers is the most important thing. Our system maintains a separate list for every label-id. Whenever a new packet arrives, its label-id is extracted, and the corresponding packet is placed in that label-id list at the appropriate position according to its sequence number. The algorithms for packet generation, sender and receiver processes are provide in Algorithm~\ref{pkt-gen}, \ref{sender}, and~\ref{receiver} respectively.  A push-based approach causes packet losses and network delays. It may lead to an inconsistent view of the board among the peers if the replay at a receiver happens without handling of delayed packets.  For handling delayed packets, at first each packet is placed at its correct relative position among other packets and later the repainting is done using the label-id and the sequence number carried in the packet. It is not a compute-intensive operation. The repainting is also very quick, as most computers can perform more than $10^8$ operations per sec. Furthermore, repainting performed only for one single page of the canvas at a time, not for the entire canvas. Therefore, repainting is quite fast.
\begin{algorithm}
  \caption{Packet generation algorithm.}
    \label{pkt-gen}
    \While{BufferNotFull OR MouseReleaseEventNotHappened} {
        Create data packet;
        SendQueue.put(packet);
    }
  
\end{algorithm}

\begin{algorithm}
 \caption{Sender's algorithm.}
    \label{sender}
    packet = SendQueue.get()\;
    \For{ nodes in neighbor} {
        Send to all except from whom it is received;
    }
   
\end{algorithm}

\begin{algorithm}
 \caption{Receiver's algorithm.}
    \label{receiver}
    receive data packet\;
    extract packet-id\;
    \If{seen[packet-id]==True} {
       continue;
   }
   \Else{
        seen[packet-id]=True\;
        SendQueue.put(packet)\;
        extract client-id, sequence no., label-id, packet length and data\;
        \If{ Label[label-id].list.last-sequence no. $\leq$ sequence\_no} {
            Label[label-id].list.append(sequence\_no.)\;
            last-sequence\_no = sequence\_no\;
            draw(data)\;
        }
        \Else {
            insert (sequence no., data) at appropriate position in Label[label-id].list\;
            repaint page with label-id from that sequence\_number\;
        }       
    } 
   
\end{algorithm}

A peer joining late may get an inconsistent view of the shareable board.  We used a combination of push-pull to solving this problem.  Initially, a comparison is done with label-id and if a conflict arises then another comparison is performed with the sequence numbers. If the current label-id or the sequence number is greater than zero, then it is safe to assume that data propagation has already started. Hence, the late joining peer explicitly requests (pulls) for the previous data from one of the neighbors. The number of operations performed on each page is stored with its corresponding page number of the canvas.  A page may perhaps be left blank intentionally. In other words, the number of operations performed on a page may be zero or a positive number. In a pull request, a user first requests for the data of the page in the current view. It can be extracted with the help of label-id.  Therefore, a pull request must include label-id — the sender replies with the number of operations performed on that page.  The receiver side then verifies the reply. If the number of operations performed on that page is zero, then, no further request is made for that label-id. Otherwise, a request for remaining operations would be made. These operations are requested one at a time in a pull-based manner until the latest sequence number is reached. When the late joiner receives all the packets only then his/her session gets activated. 

An explicit deferred joining process is necessary to avoid the inconsistent intervention of a late joiner. To understand why it happens, consider a user who joins after half an hour of the start of a session. Suppose the user immediately starts some operations on the first page of the canvas when the user's device is still in the process of receiving data from other peers. Such an intervention by the user leads to an overwriting of the previous data. Therefore, we defer the activation time of the late joining peer until the data synching is complete. 
Another solution could be to enable the group-undo feature in the system, i.e., every undo command gets propagated to the entire group, and required changes will be done on every peer's canvas. Even if the user joins the system before the sync is complete and starts scribbling, we could use a group-undo command to undo those operations. The inclusion of the group-undo feature is not available in the current work.

\section{De Bruijn Overlay for P2P File Sharing and Searching}
 
\label{sec:p2p-overlay}

The proposed P2P file sharing and searching system incorporates many additional features. The most important among these features is annotations of audio, video and PDF files.  Therefore, it is a concomitant component to P2P-PS front-end platform. It helps in reflective learning and creating mash-up presentation using stored repositories. The file sharing system is implemented using de Bruijn graph overlays.

\subsection{De Bruijn Graphs}
\begin{figure}
\centering
    \includegraphics[scale=0.3]{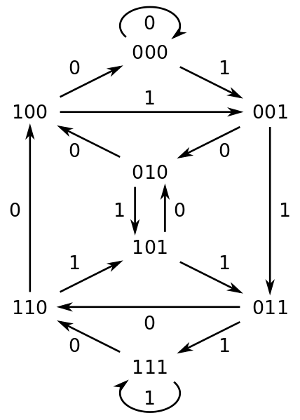}
    \caption{An example of de Bruin graph $B(2, 3)$}
    \label{fig:deBruijn}
\end{figure}

A de Bruijn graph is a labeled directed multigraph with fixed out-degree $K$. Every node of the graph has an $ID$ or a label of fixed length. Let the length of each $ID$ be $D$ and alphabet $\Sigma$ of size be $K$. De Bruijn graph $B(K, D)$ of $N=K^D$ nodes can be constructed as follows. Each node is connected by an outgoing edge to $K$ other nodes. There is an outgoing edge from a node $A$ to a node $B$ if $ID_B$ can be created by applying a left shift to the label $ID_A$ and appending one symbol from alphabet of size $K$ to the rightmost position, i.e., $ID_B = ID_A[1:]+a$ where $a\in\Sigma$. If labels are considered as base $K$ numbers, then outgoing edges will be to all the nodes with label equal to $(ID_A * K + k)(mod N)$ where $0\le k\le K-1$. Figure~\ref{fig:deBruijn} illustrates an example of a de Bruijn graph $B(2,3)$.

\begin{figure}
    \includegraphics[scale=0.4]{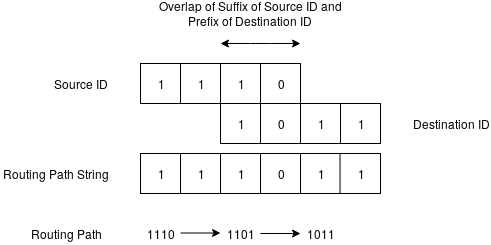}
    \caption{An example of routing in de Bruin graph}
    \label{fig:deBruijnRouting}
\end{figure}

Routing in a de Bruijn graph is specified in the form of a string. Let $S$ and $D$ denote the source and the destination nodes with labels $ID_S$ and $ID_D$ respectively. Then the routing path from $S$ to $D$ is obtained by looking at the maximum overlap between suffix of $ID_S$ and prefix of $ID_D$, and calculated as $ID_S$ appended by the non-overlapped part of $ID_D$. For concreteness, assume that a look-up for destination node 1011 is initiated by a source node with $ID = 1110$. The maximum overlap of the suffix of 1110 and prefix of 1011 is 10 as shown in Fig~\ref{fig:deBruijnRouting}. Hence, the routing path would be specified by string 111011, i.e., 1110 appended by the non-overlapped part 11 of $ID_D$. 
The next hop $v$ is derived from a current hop $u$ as follows:
\begin{enumerate}
    \item $v$ is a neighbor of $u$ in the input de Bruijn graph.
    \item The length of longest suffix of $v$ that is same as prefix of $D$ is 1 more than the length of the longest suffix of $u$ that is equal to a prefix of $D$.
\end{enumerate}  
This routing scheme is known as substring routing.

The structure and the routing method indicate that the diameter of de Bruijn graph is $D = \log_{K}{N}$. De Bruijn graphs have low clustering and exhibit $(K-1)$-node-connectivity~\cite{loguinov2003graph}. $K$-node-connectivity is not possible due to self-loops on $K$ nodes with IDs of the form $\alpha\alpha\ldots\alpha$ for $\alpha=0$ to $K-1$. The nodes can be linked together to form a ring which makes the graph $K$-regular and also achieves $K$-node-connectivity~\cite{du2002connectivity}. $K$-node-connectivity makes the graph more resilient to fault-tolerance, as even the failure of any $(K-1)$ nodes cannot disconnect the graph and diameter remains at most $D+1$.  Loguinov et al~\cite{loguinov2003graph} showed that the expected congestion in de Bruijn graph is much less than the other counterparts under a similar rate of load, due to larger bisection width of the graph.

\subsection{Comparison with Other Graphs}
De Bruijn graph possesses better asymptotic degree-diameter properties, compared to some of the widely used DHTs Chord~\cite{stoica2003chord}, Trie~\cite{freedman2002efficient}, CAN~\cite{ratnasamy2001scalable}, Pastry~\cite{rowstron2001pastry} and Butterfly~\cite{malkhi2002viceroy}.  Tables~\ref{tab:degreediameter} and~\ref{tab:diameterforN} provide a summary of comparisons in terms of degree and diameters from the analysis made by Loguinov et~al.~\cite{loguinov2003graph}. The blank cells in Table~\ref{tab:diameterforN} indicate that corresponding node degrees are not supported.
\begin{table}[!htb]
 \caption{Asymptotic degree-diameter properties.}
    \label{tab:degreediameter}
    \begin{center}
     \begin{tabular}{||c | c | c||} 
     \hline
     \bf{Graph} & \bf{Degree} & \bf{Diameter D}\\ [0.5ex] 
     \hline\hline
     de Bruijn & $K$ & $\log_K{N}$ \\ 
     \hline
     Trie & $K+1$ & $2\log_K{N}$ \\
     \hline
     Chord & $\log_2{N}$ & $\log_2{N}$ \\
     \hline
     CAN & $2d$ & $d/2 N^{1/d}$ \\
     \hline
     Pastry & $(b-1)\log_b{N}$ & $\log_b{N}$\\ 
     \hline
     Butterfly & $K$ & $2\log_K{N} (1-o(1))$\\[1ex] 
     \hline
    \end{tabular}
    \end{center}
\end{table}
\begin{table}[!htb]
 \caption{Graph diameter for $N=10^6$ nodes}
    \label{tab:diameterforN}
    \begin{center}
     \begin{tabular}{||c | c | c | c | c | c | c||} 
     \hline
     \bf{k} & \bf{de Bruijn} & \bf{Trie} & \bf{Chord} & \bf{CAN} & \bf{Pastry} & \bf{Butterfly}\\

     \hline\hline
    2 & 20 & – & – & huge & – & 31 \\
    \hline
    3 & 13 & 40 & – & – & – &  20 \\
    \hline
    4 & 10 & 26 & – & 1,000 & – & 16 \\
    \hline
    10 & 6 & 13 & – & 40 & – & 10 \\
    \hline
    20 & 5 & 10 & 20 & 20 & 20 & 8\\
    \hline
    50 & 4 & 8 & – & – & 7 & 7\\
    \hline
    100 & 3 & 6 & – & – & 5 & 5\\
     \hline
    \end{tabular}
    \end{center}
\end{table}

\begin{table}[h!]
\caption{Average distance between pair of nodes for $N=10^6$}
    \label{tab:avgdist}
    \begin{center}
     \begin{tabular}{||c | c | c ||} 
     \hline
     \bf{K} & \bf{Moore Graph} & \bf{de Bruijn}\\

     \hline\hline
    2 & 17.9 & 18.3 \\
    \hline
    3 & 11.7 & 11.9\\
    \hline
    4 & 9.4 & 9.5\\
    \hline
    10 & 5.8 & 5.9\\
    \hline
    20 & 4.5 & 4.6\\
    \hline
    50 & 3.5 & 3.5\\
    \hline
    100 & 2.98 & 2.98\\
     \hline
    \end{tabular}
    
    \end{center}
\end{table}
Table~\ref{tab:avgdist} compares the average distance between the nodes in de Bruijn graph to the optimal Moore graph~\cite{loguinov2003graph} with the same degree $K$. In a de Bruijn graph it remains very close to optimal values even for the smaller values of $K$.

\section{Overlay Network Based on De Bruijn Graphs}
\label{sec:deBruijnDHT}
Loguinov et al.~\cite{loguinov2003graph} proposed guidelines for incremental construction of a de Bruijn graph. However, the paper falls short of implementation as it does not address the problem of maintaining de Bruijn structure in presence of churning. 

For implementation, we choose the parameters $K=8$, and $D=8$ for a de Bruijn graph. It allows us around 16 million nodes inside network labeled  "00000000-77777777" in octal strings. However, the diameter of a de Bruijn graph remains O(1) (to be precise 8). The DHT overlay in our system would support efficient look-ups, and an epidemic dissemination-based protocols can be designed to infect all the nodes in just a few rounds.

\subsection{Information Structure at Each Node}

\begin{figure}
\centering
    \includegraphics[scale=0.3]{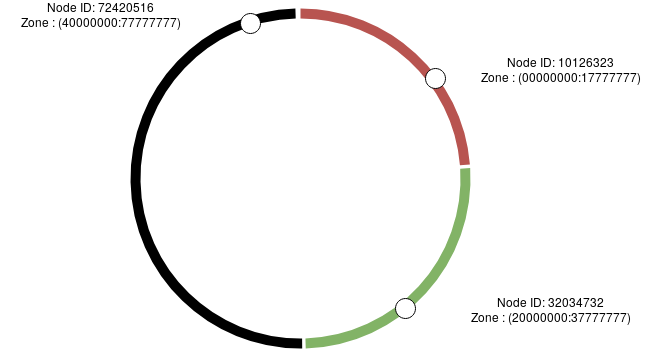}
    \caption{Three physical nodes in a de Bruijn based ring with Zone ownership.}
    \label{fig:deBruijnRing}
\end{figure}

We refer to the nodes in the underlying de Bruijn graph as {\em virtual} nodes. For maintaining the underlying  graph, a physical node in our system is responsible for a range of {\em virtual} nodes with consecutive $ID$s. A range of a physical node in our system is referred to as its zone. Initially, when a single physical node joins the DHT, it is responsible for all virtual nodes from $00000000$ to $77777777$. The virtual ID-space may be visualized as a ring, where each physical node is responsible for only a arc of the ring. Figure~\ref{fig:deBruijnRing} illustrates an example of three physical nodes responsible for three different zones.
The structure is similar to Chords~\cite{stoica2003chord}. However, unlike the Chord overlay a physical node responsible for an ID space arc may be located somewhere randomly within it. A node $A$ has an outgoing edge to another node $B$, if there is at least one virtual node in $A$'s zone having an outgoing edge to a virtual node in $B$'s zone. Each node keeps a list of its outgoing and incoming edges. Each node in these lists knows its address and zone ID. The details of the structure maintained at each node is given in Table~\ref{tab:structure}.
\begin{table*}[h!]
  \caption{Structure maintained at each node}
    \label{tab:structure}
    \begin{center}
     \begin{tabular}{||c | m{20em} ||} 
     \hline
     \bf{Item} & \bf{Information}\\
     \hline\hline
     ID & ID of node is a label lying in the Zone\\
     \hline
    Zone & A range of virtual IDs for which node is responsible \\
    \hline
    External Address & Address on which others should contact\\
    \hline
    Outgoing Edges & List of nodes connected by outgoing edge, along with info about their Zone and Address\\
    
    \hline
    Incoming Edges & List of nodes connected by an incoming edge, along with info about their Zone and Address\\
    \hline
    \end{tabular}
  
    \end{center}
\end{table*}
\subsection{Joining of Nodes}
Suppose $A$ tries to join. It selects a random ID and forwards a join request to identify the owner of the zone in which the chosen random ID falls. For convenience in the description we use the following convention:
\begin{itemize}
    \item Any intermediate node receiving the join request initiated by $A$ is referred to as $B$ 
    \item The node whose zone contains the random ID chosen by $A$, is  node $C$.
\end{itemize}
The problem of joining is split into three parts according to actions of $A$, $B$ and $C$ as explained below.

\subsubsection{Actions of $A$}
Node $A$ requests the rendezvous server (whose public IP address is known to all) to provide the external address (IP address) of $A$ and those in a list of random peers. $A$ picks one peer randomly from the list supplied by rendezvous server and sends a join request to that peer. The request contains the following:
\begin{enumerate}
    \item $A$'s external address
    \item $A$ random virtual ID from the entire ID-space i.e. $00000000$ to $77777777$.
\end{enumerate}
The join request initiated by $A$ tries to identify $C$, the owner of the random ID sent in the join request. When node $A$ sends such a request for the first time, its ID is chosen using SHA-1 hash value of its external address. But, upon retries, a random ID from the entire region is picked.


\subsubsection{Actions of $B$} 

$B$ on receiving the join request forwards it to the next node on the routing path. If a routing path is not provided in the request, $B$ will create one using $B$'s ID and destination ID given in the join request. The correct path will be sent along with the join request.

\subsubsection{Actions of $C$}
On receiving the join request from $A$ (possibly through intermediate nodes) $C$ sends the information regarding its zone ID, the incoming and the outgoing links. $C$ does not accept further Join requests until joining of $A$ is complete, or the timeout of 10 seconds occurs. 

The reply from $C$ contains a structure maintained at $C$. It includes:
\begin{itemize}
    \item Virtual node label of $C$
    \item External address of $C$
    \item Outgoing and incoming edges of $C$
\end{itemize}
$C$ waits for $A$ to complete joining process send information about $A$'s new zone. $C$ then updates its own zone as follows. 
\begin{itemize}
    \item Changes its zone
    \item Sends keys with values to be managed by $A$
    \item Notifies the neighbors about the change in zone
    \item Drops the edges destroyed due to the shrinking of its zone
\end{itemize}
\subsubsection{Action of $A$ After Receiving the Reply}
$A$ picks the half part of the zone not containing $C$'s ID and chooses a random ID (label) from the picked zone as own ID.  $A$ sends information of its ID and its zone ID to $C$, and the attachment of node $A$ in DHT overlay is complete.  $A$ needs to perform the following actions to complete the joining process:
\begin{itemize}
    \item Disseminates its ID and zone information to all the links shared by $C$
    \item Identifies the links to be dropped and checks if there is any new link to $C$. Then makes the corresponding changes to the incoming and the outgoing edges list.
\end{itemize}
$A$ willingly accepts load shared by $C$. For the interest of space the algorithm has not been included here.  The interested readers can review the algorithm appearing in appendices~A and~B.

\subsubsection{Join Example}
\begin{figure}
\centering
\subfigure[First node.]{ \label{fig:deBruijnJoin1}\includegraphics[scale=0.5]{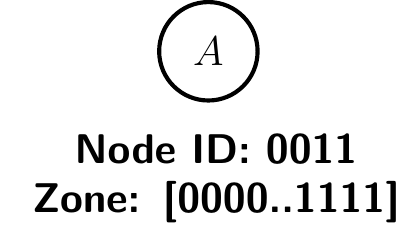}}
\subfigure[First node.]{ \label{fig:deBruijnJoin2}\includegraphics[scale=0.5]{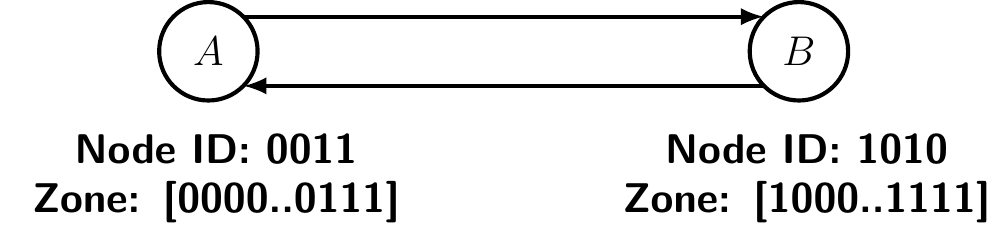}}
\caption{Joining of the first pair of nodes in the system.}
\end{figure}

For keeping the examples small, we show the Node Join process in de Bruijn graph with $K=2$ and $D=4$. Figures~\ref{fig:deBruijnJoin1} and~\ref{fig:deBruijnJoin2} illustrate joining process of the first two 
nodes in the system.

\begin{figure}
\centering
\subfigure[$A\rightarrow B$ link gets removed.]{ \label{fig:deBruijnJoin3}
\includegraphics[scale=0.5]{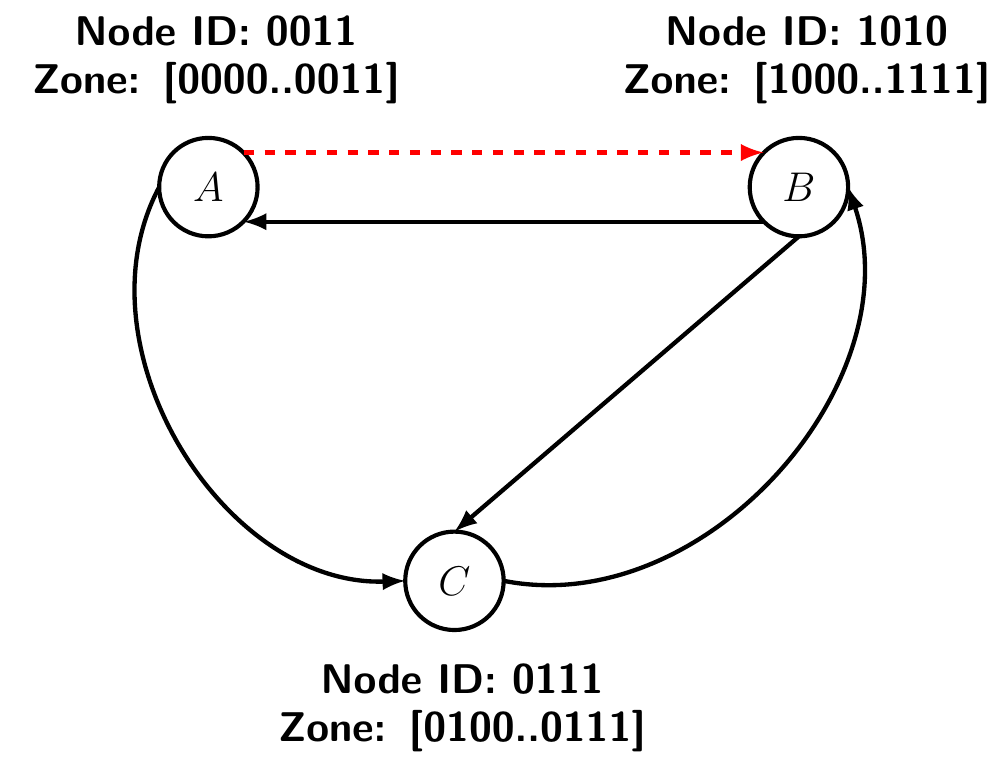}}
\subfigure[$C\rightarrow D$ \& $D\rightarrow B$ links get added.]{ 
\label{fig:deBruijnJoin4}\includegraphics[scale=0.5]{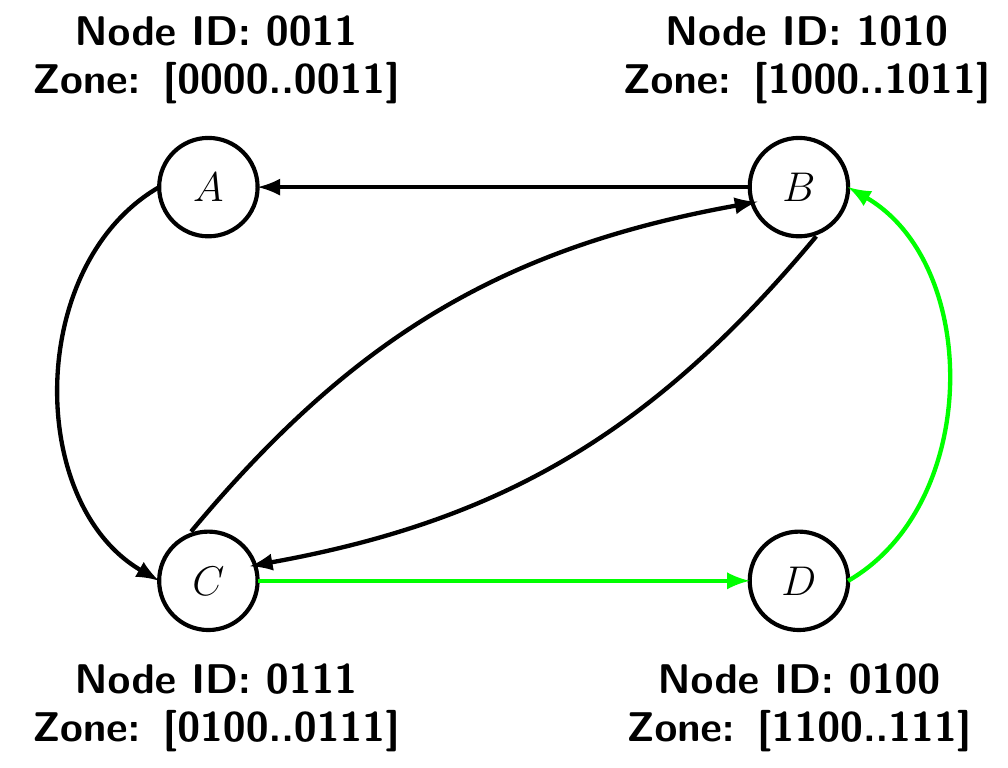}}
\caption{Joining of the next pair of nodes.}
\end{figure}

The process of joining of the third node is shown in Fig.~\ref{fig:deBruijnJoin3}. Notice that the joining of the third node requires removal of link $A\longrightarrow B$ (dashed red line). 
For the joining of the fourth node two new links are to be inserted as shown by green lines in Fig.~\ref{fig:deBruijnJoin4}.

\subsection{Leaving of Nodes}
Consider the leaving of a node $A$ from the system. A node $C$ is identified who can merge the zone of $A$ into its own Zone.  There are two part to the leave process, as leaving node is aware of its both the predecessor and the successor in the DHT overlay. 

\subsubsection{Actions of $A$}
$A$ identifies a successor zone and a predecessor zone. The first virtual ID belonging to the successor zone equal to $A$'s Zone.endID + 1. Similarly, the last virtual ID belonging to the predecessor Zone is equal to $A$'s Zone.startID - 1. $A$ picks one of the two IDs, finds the respective owners of the zones and sends a request to leave. The leave request contains the information regarding ID, Zone, incoming and outgoing links to the chosen neighbor, say $C$. If $C$ agrees within timeout period of 5 seconds, $A$ sends the load to $C$ and the leave process is complete. Otherwise, $A$ tries the same with the other neighboring node. If the timeout occurs again, the entire procedure is repeated assuming chosen nodes were busy in other Leave procedure. 

\subsubsection{Action of $C$} 
Node $C$ receiving the request checks if it is going to leave the system. If not, it accepts the request and agrees to take over the load by merging the two zones. The merging process includes merging of the incoming and outgoing links as well. $C$ then notifies all the linked nodes regarding the update. Algorithm~\ref{alg:recdLeave} in Appendix~C gives a step-wise description of $C$'s operations.

\subsection{Zone Update}

During join and leave, nodes whose zone changes notify the nodes linked to them by outgoing or incoming edges. When such notification is received, a node adds or drops links caused by the change.

\subsection{Keep-alive and Failure detection}

Every two minutes, each node sends keep-alive message to all linked nodes. These nodes update the timestamp with the zone information corresponding to that node. If no update is received in last five minutes for a linked node, it is considered dead, and owner of the successor zone is notified to take the responsibility of orphaned zone.

\subsection{Links between Zones}

If there is an outgoing edge from a virtual node in one zone to a virtual node in another zone, then there exists an outgoing edge between one node to the other. On many occasions, zones of the neighboring nodes change. A brute force way to find an edge in the underlying graph, would slow down join and leave operations. 

Node $A$ receives zone update from another node $B$. $A$ checks if the size of its zone is larger than or equal to $N/K$. If so, then a link exists (because every possible suffix of length $D-1$ is present in $A$'s zone). Otherwise, it determines the range $[L_A, R_A]$ of the suffix of length $D-1$ lying in the  $A$ and range  $[L_B, R_B]$ of the prefix of length $D-1$ lying in $B$'s zone. If $L_A \in [L_B, R_B]$, or $R_A\in [L_B, R_B]$ then there exists an edge from node $A$ to node $B$. Similarly, there is an edge exists if $L_B$ or $R_B$ lies within $[L_A, R_A]$.

\subsubsection{Examples}
\begin{itemize}
    \item Consider two zones: node $A$'s zone ['00000000', '17777777'] and node $B$'s zone ['40000000', '77777777']. Since the size of zone of node $A$ is larger than '10000000', it will have all possible suffix of length $D-1$. Therefore, it will have link to any node in the system. Same is the case for  $B$ and hence, both have outgoing edges to each other.
    \item Consider another two zones: node $A$'s zone ['00000000', '00777777'] and node $B$'s zone ['01000000', '01777777']. The range of suffixes of length $D-1$, for node $A$ is ['0000000', '0777777']. The range of prefixes of length $D-1$, for node $B$ is ['0100000', '0177777']. There is a non-empty intersection between the two ranges, and hence an edge exists from node $A$ to node $B$. (In the underlying graph, '00100000' has an outgoing edge to '01000000'.)
    \item For the above example, range of suffixes of length $D-1$, for $B$ is ['1000000', '1777777']. The range of prefixes of length $D-1$, for $A$ is ['0000000', '0077777']. The intersection between two ranges is empty, and hence no edge can exist from  $B$ to $A$.
\end{itemize}

\section{Sharing, Searching, and Global Indexing}
\label{sec:indexing-and-searching}

The implementation supports two types of searches (i) by ID/key, and (ii) by keywords. A user can share one or more directories which contain the file(s). By default, at least one directory is shared which may be empty. The said directory is located in the Downloads folder of the user's system. Anything downloaded from the P2P system is stored in this directory and is automatically shared. 

\subsection{Hashing and Key Generation}
We have used a 160-bit SHA-1 value for the contents of the file and also use it as a key. Since, each node has 24-bit label, the first 24 bits of a key are used for the routing purpose. For example, first 24 bits of the SHA-1 hash value "2fd4e1c67a2d28fced849ee1bb76e7391b93eb12" is "2fd4e1" which is equal to "13752341" in octal.

\subsection{Pre-processing of Files}
At the start of the application and after every 30 minutes, the directories are scanned for any new or modified files. The new and modified files are queued for processing. SHA1 of a file is computed, and it is stored in the local database. Any deleted or renamed file is also identified. For deleted files, the entries related to them are deleted from the local database.  A set of important keywords are associated with file to improve the chances of finding a document or file in the DHT. We provide support for extracting text from PDF and video files that come along with the subtitles. If a PDF is created using scanned copies, the text is extracted using an OCR tool. Once the text is extracted,
TF-IDF scheme is used to identify top 100 keywords for a file using its content. In this scheme, the file is considered to be a set of documents of 1000 words each. In the TF-IDF scheme, a term is assigned a higher score if it has larger term frequency other terms, but appears in a less number of the documents. The scoring scheme is, therefore, based on Inverse Document Frequency. These keywords are stored in the local database along with file's modification times, which can be used to avoid processing of the file again.  For all keywords, SHA-1 is also computed to obtain the keys. The users can also manually add up to 10 keywords. 

\subsection{Global Indexing of Files}

A Key is a globally unique ID of an object (e.g., SHA-1 value), and the Value (the name of the object), the size of the object, the address of the object and the timestamp of the last refresh. The keywords are stored as meta data and maintained in the local database at the external nodes of an overlay. The purpose is to enable keyword based search for related document and media files. We need two basic operations, namely, Put and Get.

For each file, MultiKeyPut is used to insert key-value pairs of all keywords in the system. The keywords also include words in the file name. The associated value stores information such as SHA-1 hash of file, size of the file, name of the file and address of the file. After every 30 minutes, MultiKeyPut is called for keywords of each file, thus refreshing the timestamp for keys in corresponding nodes. A node will check for the keys' timestamp every 10 minutes, if a particular node has not updated a key in last 60 minutes, corresponding key-value pair is considered to be not available and dropped from the local store. The delay of 60 minutes is kept considering a lost update in the network.

\subsection{File Searching}

Searching operation is performed one at a time. A user can cancel the search at anytime. Since, there can be delayed replies, a unique search ID is associated with the Get requests to distinguish the results. Any node returning result/s must provide search ID along with the result. The packets received for the current search ID are kept, and all other are discarded.

\subsubsection{Search by keywords}
A query is broken down into keywords, and search is conducted with MultiKeyGet procedure. For any reply that comes while search is active, the ranking of the results may change. The file with more number of query keywords in it, gets higher rank. The files with same number of query keywords is distinguished by the replication/popularity in the system. The more popular result gets a higher ranking than the others. 

\subsubsection{Search by file ID}
A user might receive the ID or the key of a file from another user via some communication channel. If user wishes to download file with given ID, he/she can enter it in Get File dialog which user Get method. On receiving the results, a download request is initiated. 

\subsubsection{File Download}
When results are fetched using the above search methods, a user can choose to download a file. If chosen file is available with multiple peers, multi-threaded approach for downloading is used. Different chunks are requested from multiple peers, and they are written to the file as they are downloaded.

\section{Annotations}
\label{sec:annotations}
Most PDF viewers have tools to highlight a portion of the text. It is useful for the revision of the text or skimming through the text. But, any editing of the content changes hash value of a file. Modification of hash value of a file is undesirable, as the file in P2P learning environment is archival in nature. Therefore, the features should be provided without editing the file itself. Furthermore, the users may also want to see portions that are highlighted by others. Some other users may wish to tag some text or provide a hyperlink to some other external resource. All these operations are grouped as peer learning activities without any change in the contents.  We defined a format for annotations for the purpose. The annotations are stored separately in the local database, and synchronized among peers. 

\subsection{Annotation Format}
We propose a general annotation format. This general format is a template for specific formats. The  following information is stored in an annotation:
\begin{itemize}
\item  Annotation ID: A unique ID to identify an annotation.
\item File ID: A unique ID for file. SHA-1 value is used.
\item Time stamp: Time of creation of the annotation.
\item Author: Identity of the creator.
\item Text: Associated text, if any. 
\item Properties: Specific properties of the annotation.
\end{itemize}
The "Properties" field stores the properties for specific types of annotations. The text field may be in the form of a hypertext, as the support for links and images is provided.

\subsection{PDF Annotations}
Highlighting in the PDF documents can be done by either selecting text or selecting a rectangle. We define properties of each type separately. 
\subsubsection{Text Selection}
Text selection provides the flexibility of selection of a specific word or a set of words, or a set of lines in a particular page. If selection spans multiple pages, the user should select words or lines on each page separately. The user may also provide a comment or add some text related to a highlighted area. The additional text/comments are stored with the annotation. Since each annotation of this type is designed for only one continuous region of a text, we define following properties for an annotation.
\begin{itemize}
\item PageNumber: Denotes the page number on which annotation is present.
\item FirstWord: Denotes the index of the first selected word
\item LastWord: Denotes the index of the last selected word
\end{itemize}

\subsubsection{Rectangular Selection}
Rectangular region selections are generally helpful in inserting annotation related to diagrams, or non-textual information. A user can choose one rectangular region of a page and highlight the portion. Following properties define each rectangular region:
\begin{itemize}
\item PageNumber: Denotes a page number on which the annotation is present.
\item TopLeft: Denotes co-ordinates of the top-left corner of the selected rectangle with respect to the the top-left corner of a page
\item BottomRight: Denotes the co-ordinates of the bottom-right corner of a selected rectangle with respect to the top-left corner of a page. 
\end{itemize}

\subsection{Audio/Video Annotations}
Video annotations are similar to Audio annotations. Therefore,  we do not distinguish between the two. A user can select a duration of the video and tag it. The two end-points of a duration are rounded to the nearest integers and at least 5 seconds of the duration is selected. The properties for such a selection are as follow:
\begin{itemize}
\item StartTime: The time at the start of duration.
\item EndTime: The time at the end of duration.
\end{itemize}

\section{Discussion Forum and Announcements}
\label{chap:p2p-forums}
As discussed in the previous section, our system provides a way to create an annotation and save it. A user can use them for personal use. He/she can also share them with others to spread useful information or ask a doubt regarding the material. An annotation is disseminated to other peers and can be discussed among the participating peers. So, we integrated a discussion forum along with annotation announcement. 

When a user creates an annotation, he/she has an option to share it with others. If the user chooses to share it, the annotation is converted into a post and gets disseminated among the peers. Other peers can put comments on the shared post.

\subsection{Post}
A post inherits the format of the corresponding annotation, adds an additional field for Title. A post also has a constraint of minimum number of 100 characters in the Text field. The maximum number of characters is limited to 1600.
\begin{itemize}
\item ID of Post: A unique ID to identify a post
\item File ID: A unique ID for file. SHA-1 value is used
\item Time stamp: The time of creation of annotation
\item Author: The identity of the creator
\item Text: Associated text
\item Title: A short description of the post
\item Properties: Specific properties of the PDF or Audio/Video annotations are stored.
\end{itemize}
When an annotation is converted into a post, Text field of Annotation is copied into the Text field of the Post. The properties field of Post supports multiple annotations. Multiple annotations are stored in a list, serialized into a string and then stored in Properties field. If multiple annotations are to be attached to a post, File ID is set to zero. 

\subsection{Comment}
The comments on a post follow similar format as the post. But, instead of Title field, it has ReplyTo field.
\begin{itemize}
\item Comment ID : A unique ID to identify a comment
\item File ID: A unique ID for file, SHA-1 value is used
\item Time stamp: The time of creation of the annotation
\item Author: The identity of the creator
\item Text: Associated text
\item ReplyTo: ID of post or comment to which this is a reply
\item Properties: Specific properties of the PDF or Audio/Video annotations are stored.
\end{itemize}

\subsection{Synchronization of Posts and Comments}

Posts and comments are synchronized with the help of two protocols.
\begin{itemize}
    \item An epidemic dissemination based protocol is used to spread recently published posts and comments.
    \item  Since, a user won't be active at all times there will be missed updates. For such cases, reconciliation (an anti-entropy session) has to be performed at regular intervals with the neighbors.
\end{itemize} 
The consistency requirements of the discussion platform is weaker because a post once posted, cannot be updated. For a particular post or a comment, only one person is responsible. Posts and comments need not be delivered immediately. The happened-before relation of the comments and the posts is maintained due to their hierarchical structure. The causal ordering of other messages can only be provided through time-stamps associated with them. We assume that clocks of the machines are synchronized with NTP servers.

\subsubsection{Epidemic dissemination}
Every node receives posts and comments through their neighbors. Our application periodically checks if any new post or comment arrived since the last check. If there are new posts or comments, they are sent to all the neighbors except the ones from which they were received. This ensures the delivery of messages to all. The period of such check is set to one minute. The diameter of our system is $D=8$, therefore, any new message gets delivered to the entire system in less than ten minutes. The interval time can be set to zero, and in that case, messages will be delivered to others instantly.

\subsubsection{Reconciliation}
Reconciliation procedure executes at regular intervals of thirty minutes. It begins by randomly picking a neighbor at the start of an interval. A predefined time of seven days is chosen. When posts from last seven days are sorted by time, the first post's timestamp is picked, and shared with the neighbor. On receiving the reconciliation request. The neighbor chooses all the posts that started on or after the given timestamp, and sorts them according to timestamps. A list of IDs of these posts or comments are created. The list is divided into chunks of 256 entries. A hash value is calculated for each chunk, based on the post/comment IDs. These hash values are shared with the initiator. The initiator also calculates the hash values in the similar fashion, and the hash values are compared. If the hash value is different, then the initiator asks a neighbor to share the IDs in the particular chunk. It then gets the list of IDs in the chunk from the neighbor. Both of them make changes to the corresponding lists by adding missing IDs and recalculate hash values for that chunk. New list of hash values from the mismatched chunk is sent to the initiator. When the last chunk's hash matches reconciliation is over. Most of the time, only last few chunks would differ, and the number of message exchanges is low. 

\subsection{Announcements}
Announcements are similar to posts in the discussion forum. They are useful for posting information about locations of newly added documents. The sharing of announcements is carried out in a similar fashion. They are forwarded to other peers as soon as they are received. On the front-end, announcements appear separately from other posts.

\section{Results of Simulation}

Emulab provides an environment for experiments over test beds consisting of large-scale distributed network. We deployed scripts for the experiments using Emulab portal to acquire both physically distributed and purely simulated nodes~\cite{emulab2008}. It fitted the kind of experiments we wanted to run for establishing efficacy of our approach to build a low latency file sharing, searching and inline annotations framework ideally suitable for E-learning system using a P2P organization. With respect to whiteboard sharing our primary motivation was to find out how a P2P learning system may work in a LAN environment and support up to a maximum of 250 nodes. However, the simulation experiments were carried out on up to 1000 nodes.

\subsection{Live Streaming and Whiteboard}

The first experiment is to find out the stability of our system. 
During the process of joining as explained in section~III-A, a peer gets a list of $\log_2 n$ active peers. Running simulations on up to $1000$ nodes, we found that the choice of $\log_{2}n$ leads to a stabilization of network as the size of the overlay increases. Once the bootstrap server returns the list, the peer send 'Adopt me' request to all the active peers in the list. A peer receiving join request could accept or discard the request based on its fanout value. With simulations on up to $1000$ nodes, we found that the choice of $\log_{2}n$ leads to a stabilization of network as the size of the overlay increases. We found that maximum path increases with an increase in the number of nodes, but later it gets stabilized. As shown in figure~\ref{fig:wnchurn} from 700 to 1000 nodes, the maximum path length has stabilized to 6 units.
 \begin{figure}[t]
    \centering
\subfigure[Maximum path length.]{\label{fig:wnchurn}\includegraphics[scale=0.4]{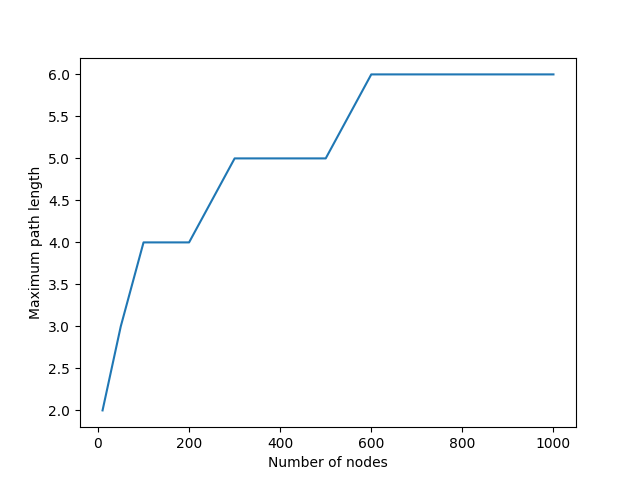}}
   \subfigure[Churning with 10\%, 20\%, and 30\% churn rates.]{\label{fig:churn}\includegraphics[scale=0.4]{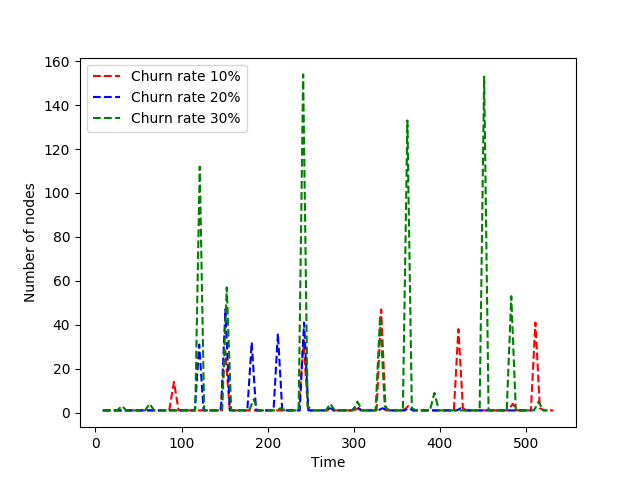}}
    \caption{The results on Emulab for stabilization of the maximum path length and churning.}
    \label{fig:wnchurn}
\end{figure}
Mesh-overlay might get disconnected in the presence of churn. A node is having $0$ in-degree implies that it has no parent and hence, it could not receive data until it finds at least one parent. So, even in the presence of churning it is important that the overlay should eventually get stabilized. It means only one node could have in-degree $1$. This node should be the source node only.
We performed on simulations on 1000 nodes for churn rate 10\%, 20\%, 30\%. The results appear in figure \ref{fig:churn}. The number of nodes counts the nodes having in-degree equal to zero. The time is measured in seconds. We observed that even with 30\% churn rate, the overlay stabilizes within 5 seconds.

Table~\ref{tab:throughput} depicts the minimum throughput value analyzed on a different number of nodes. Here, throughput is defined as the number of packets received in one second.
\begin{table}[h!]
\caption{Minimum throughput value.}
\label{tab:throughput}
\begin{center}
\begin{tabular}{ ||c|c|c|c|| } 
\hline
\bf{Number of nodes} & \bf{Throughput} \\
\hline\hline
20 & 177 \\ 
\hline
50 & 152 \\
\hline
90 &  173\\
\hline
120 & 156\\
\hline
150 & 139\\
\hline
170 & 173\\
\hline
200 & 174\\
\hline
\end{tabular}
\end{center}
\end{table}
We have obtained these results by performing experiments on Emulab. In Emulab experiments, each node receives 5000 packets from its parent nodes. As discussed in section~VI, the number of packets generated per second = 179. Hence, 5000 packets will be generated in $\frac{5000}{179} = 27.93$ seconds. Hence, we can see from Table~\ref{tab:throughput} that our results are closely related to the theoretical results.

\subsection{De Bruijn Overlay}

De Bruijn graph has diameter and out-degree equal to eight~\cite{}. At the application layer, the diameter does not exceed the value of eight. But, out-degree can vary according to size of a zone. In theory, the maximum out-degree would be less than $K \times O(\log_K{N})$ with high probability, where $N$ is the number of nodes in the system. Considering $N=800,000$ nodes, the maximum out-degree among all runs was 41. Experimenting with different values of $N$, the average out-degree was found to be 7.99. It is due to absence of self-loops. 
\begin{figure}
    \centering
    \includegraphics[scale=0.5]{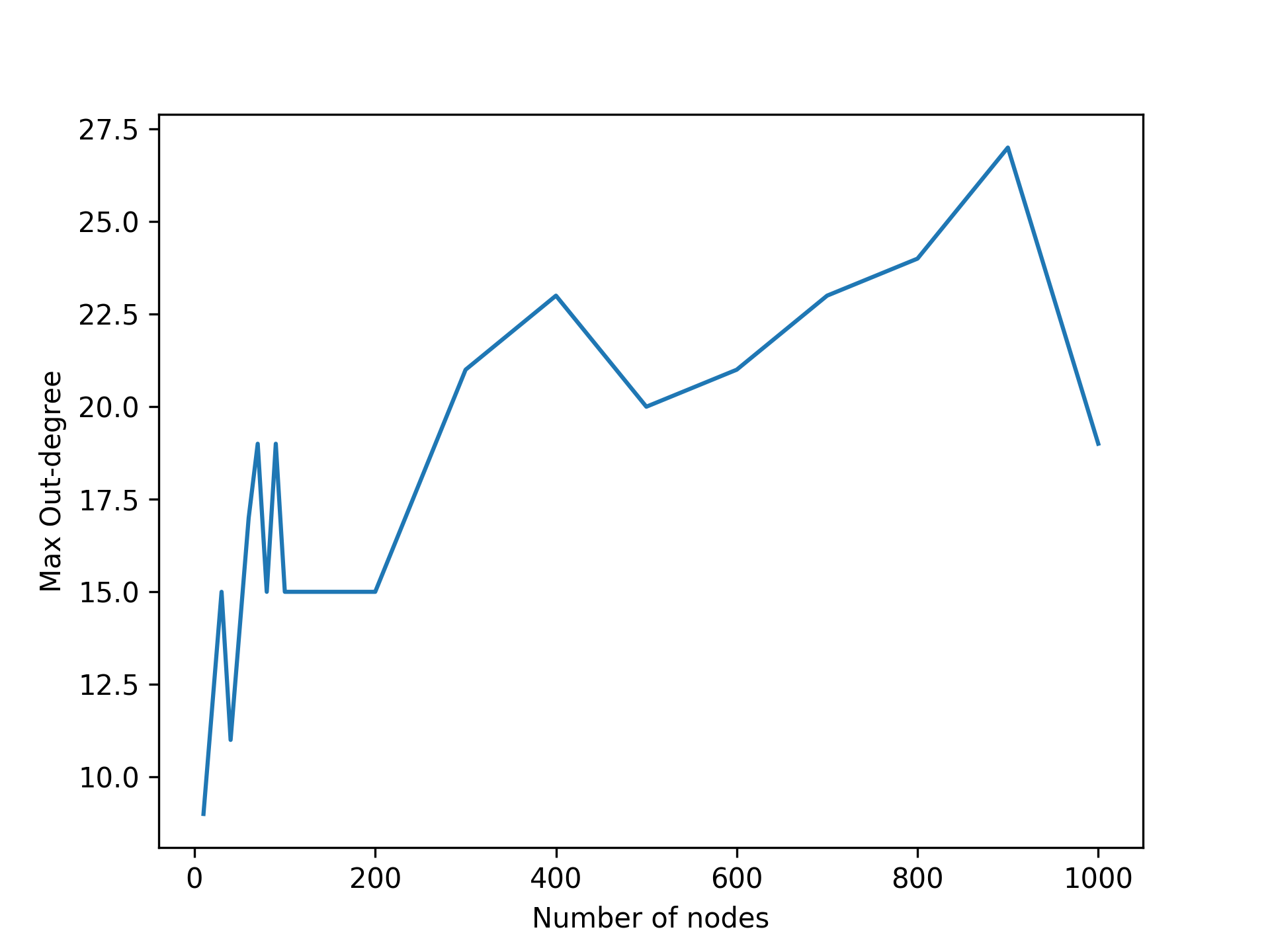}
    \caption{Maximum out-degree of nodes}
    \label{fig:upto-thousand}
\end{figure}
\begin{figure}
    \centering
    \includegraphics[scale=0.5]{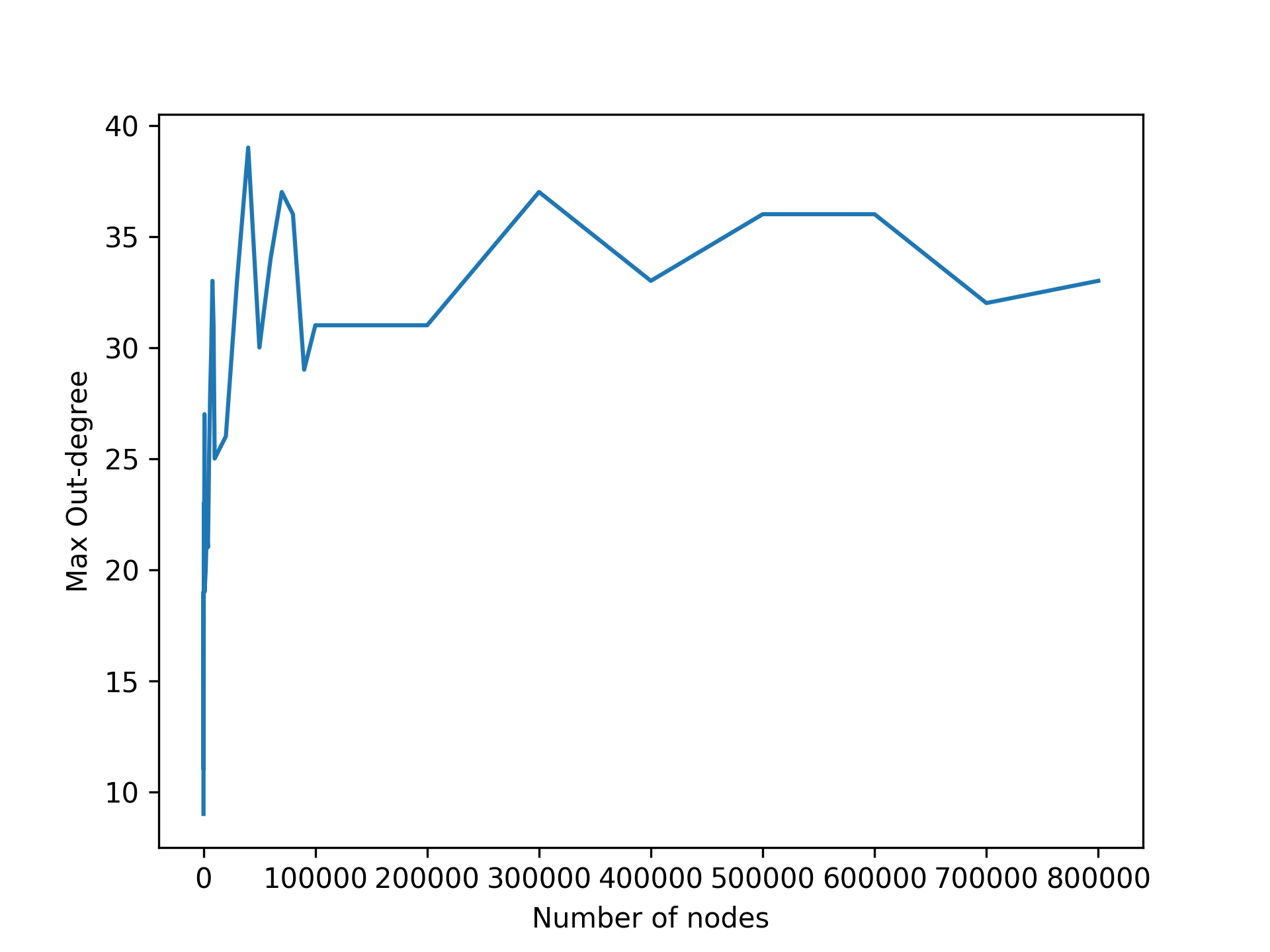}
    \caption{Maximum out-degree of nodes}
    \label{fig:upto-million}
\end{figure}
\begin{figure}
    \centering
    \includegraphics[scale=0.5]{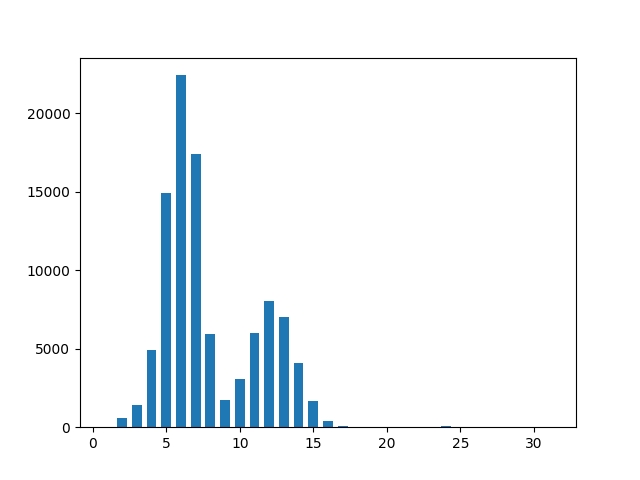}
    \caption{Distribution graph of out-degree of nodes}
    \label{fig:key-distro}
\end{figure}

The graph in Fig~\ref{fig:upto-thousand} and Fig~\ref{fig:upto-million} show that the median values of the maximum out-degree of a node. The simulation results match with the theoretical results. 
The graph shows that the median value for the maximum out-degree is 31 when $N=100000$. The distribution of out-degree for the same is shown in Fig~\ref{fig:key-distro}. It shows that very few nodes have degree $> 2\times K$. Just 380 out of 100000 nodes have degree $> 16$, which is only $0.38\%$ of the total. None of the nodes have out-degree $> K \times \log\_K{100000}$. In an equivalent Chord implementation, the average out-degree is $O(\log N)$ which is $>$ 20. Our experiments also determined that, minimum in-degree=7, and the maximum in-degree=8. 

We also carried out experiments on Emulab to find out the maximum latency and the success rate of look-ups. In the experiment, the nodes were connected to each other in a LAN environment. Every node randomly picked twenty-five words from the list of three thousand words available to everyone. After a delay of five minutes, the nodes sent out queries for the same set of words. The boot-up times of machines in Emulab differ up to $\pm 5$ minutes. Hence, the join procedure for de Bruijn overlay network would have involved a transfer of the load for most of the nodes. The results were calculated for the varying number of nodes as shown in Table~\ref{tab:exp1}. It indicates that none of the lookups failed, even during the dynamic joins in the set up.
\begin{table}[h!]
 \caption{Lookup latency and success rate}
    \label{tab:exp1}
    \begin{center}
     \begin{tabular}{||c | c | c | c||} 
     \hline
     \bf{No. of nodes} & \bf{Success rate} & \bf{Max Lat.(ms)} & \bf{Avg. Lat.(ms)}\\
     \hline\hline
    80 & 100\% & 50 & 16 \\
    \hline
    120 & 100\% & 70 & 20\\
    \hline
    160 & 100\% & 58 & 21\\
    \hline
    200 & 100\% & 52 & 21\\
    \hline
    \end{tabular}
   
    \end{center}
\end{table}

In another experiment, the nodes are allowed to leave the system with 10\% probability after every three minutes. Our approach achieved success rate of 99.39\% and the maximum latency of 52ms for 200 nodes. Assuming that the human tolerance limit is about 200ms, the response is pretty good

\section{Conclusion}

We have studied two different architectures to support live streaming, and based on the analysis we modified mesh architecture to support dynamic fanout leveraging spare capacity whenever available. Hence, our system could also support heterogeneous nodes satisfying the minimal bandwidth requirement as explained in section~VI. In our experience with actual sessions on live board, we found a consistent display of results. The repainting was quite efficient at all peer nodes. The buffered approach has helped to optimize the shareable whiteboard's performance. 

Our simulations on Emulab, proved that even in the presence of churning, the overlay structure for live streaming gets stabilized moderately quickly. Since the maximum path length also gets stabilized the latency is bearable in Live sessions of P2P interactions with live streaming. The Emulab experiments are important to establish that experimental throughput is close to theoretical throughput values.

As a part of future work, we plan to include more tools in our live shareable whiteboard and enable group-undo operation. Various video-coding techniques are available to neutralize the effect of packet loss. In the future, we would like to incorporate those techniques for live streaming. The back end support for file sharing, searching and inline annotation also needs to be integrated.
\balance
  \bibliographystyle{IEEEtran}
  
    \bibliography{citations}
\newpage
\onecolumn
\begin{appendices}
\section{Send Join Request Algorithm}
\begin{algorithm}
\caption{SendJoinRequest}
\label{alg:joinReq}
\SetKw{KwReq}{Require:}
\KwReq{Addresses of the rendezvous server \& its socket\;}
send connect request to the server\;
receive \& store own extAddress\;
receive ListOfPeers from the server\;
randomID $\gets$ pick a number from (0, $K^D$-1) based on SHA-1 of extAddress\;
\If{(ListOfPeers == $\Phi$)}{
     MyZone $\gets$ entire range of virtual nodes\;
     MyID $\gets$ randomID\;
     MyOutEdges, MyInEdges $\gets$ $\Phi$ \;
 }
\Else{
    joined $\gets$ FALSE\;
    \While{(!joined)} {
        randomPeer $\gets$ select a random peer from ListOfPeers and delete it\;
        send JoinRequest(randomID, own extAddress) to randomPeer\;
        wait for up to 10 seconds for message from owner(randomID)\; 
        \If{(timeout)}{
            \If{(ListOfPeers == $\Phi$)} {
                 receive new peer addresses from rendezvous server\;
                 update ListOfPeers with newly received addresses\;
                 randomID $\gets$ randomNumber(0, $K^D$-1)\;
            }
           \Else {
           \If{(Zone contains only one virtual ID)}{
               randomID $\gets$ randomNumber(0, $K^D$-1)\;
               continue and try another node\;
            }
            \Else {
                 recv (ownerID, Zone, OutEdges and InEdges)\;
                 split the received Zone into two parts Zone.part1, Zone.part2\;
                 if {(ownerID $\in$ Zone.part1)} {
                      discard Zone.part1\;
                      MyZone $\gets$ Zone.part2\;
                      }
                      else {
                           Discard Zone.part2\;
                           MyZone $\gets$ Zone.part1\;
                      }
                 MyID $\gets$ randomID(MyZone)\;
                 MyOutEdges, MyInEdges $\gets$ received OutEdge, InEdges\;
                 send notifyJoin(MyZone,MyID) to previous zone owner \;
                 send notifyJoin(MyZone,MyID,extAddress) to all neighbors\;
                 check if any new edge exists between previous owner's new zone and my zone\;
                 drop edges (with any of the neighbors) that no longer exists\;
                 
                 // For adding to list of peers\;
                 notify rendezvous server\;
            }
        }
    }
}
}

\end{algorithm}

\newpage
\section{Received Join Request Algorithm}
\begin{algorithm}
\caption{ReceivedJoinRequest}
\label{alg:processingJoin}
\SetKw{KwReq}{Require:} 
\KwReq{Join request packet JOIN(ID, JoinExtAddr)\;}
\If{(ID $\in$ MyZon)}{
    send MyZone, MyID, MyOutEdge, MyInEdges to JoinExtAddr\;
    queue new joins until this join completes or timeOut==10s\;
    receive ID and Zone information of the new node\;
    split MyZone into half and discard the new node's Zone\;
    identify and insert in appropriate lists any new edges with the newNode's Zone and MyZone\;
    notify all neighbors about update of Zone by sending MyZone and MyID\;
    drop edges (with any of the neighbors) that no longer exists\;
    share existing Key, Value pairs in lying new node's Zone with the new node\;
}
\Else { // Receiver is a forwarder\;
    identify the next node on routing path to which request is to be forwarded\;
    send JoinRequest(ID, JoinExtAddr) to next node\;
}
\end{algorithm}

\section{Send Leave Request}
\begin{algorithm}
\caption{SendLeaveRequest}
\label{alg:leaveReq}
\SetKw{KwReq}{Require:}
\KwReq{The address of the rendezvous server, socket\;}
left $\gets$ FALSE\;
\While{(!left)} {
    \If{(MyZone covers entire ID-space)}{
        inform the server that all other nodes have left\;
        left $\gets$ TRUE\;
    }
    \Else{
        list $\gets$ [MyZone.startID-1, MyZone.endID + 1]\;
        // succID is the ID of the successor\; 
        succID $\gets$ randomID(list)\; 
        \If{succID lies with one of the neighbors} {
            next\_node $\gets$ address of the neightbor containing the succID\;
            }
        \Else{
            create a routing path from MyID to succID\;
            next\_node $\gets$ address of the next node from the routing path\;
        }
        send LeaveRequest(ID, Zone, OutEdges, InEdges, extAddress) to next\_node\;
        wait for reply up to 5 seconds\;
        \If{(Reply == "Yes")} {
            send confirmation and keys in MyZone to successor\;
            inform the rendezvous server, that leave is complete\;
            left $\gets$ TRUE\;
        }
    }
}

\end{algorithm}

\newpage
\section{Received Leave Request}
\begin{algorithm}
\caption{ReceivedLeaveRequest}
\label{alg:recdLeave}
\SetKw{KwReq}{Require:}

\KwReq{Leave request packet LEAVE(ID, Zone, OutEdges, InEdges, LeaveExtAddr)\;}
// LeaveExtAddr is the external address of leaving node\;
\If{(ID $\in$ MyZone)} {
    \If{(not leaving the system)} {
        send "Yes" to LeaveExtAddress\;
        wait for confirmation up to 5 seconds\;
        \If{(confirmed)} {
            merge Zone, OutEdges and InEdges\;
            receive the load\;
        }
    }
    \Else { 
        // Leaving self\;
        send "No" to LeaveExtAddress\;
    }
}
\Else {
    identify the next node on path to which the request should be forwarded\;
    send LEAVE(ID, Zone, OutEdges, InEdges, LeaveExtAddr) to the next node\;
}
\end{algorithm}

\end{appendices}
\end{document}